\documentclass[usenatbib]{mn2e}
\usepackage{graphicx}
\usepackage[english]{babel}
\title[SDSS~J124058.03$-$015919.2: A New AM~CVn Star]{SDSS~J124058.03$-$015919.2: A new AM~CVn star with a 37-minute orbital period}
\author[G.\,H.\,A. Roelofs et al.]{G.\,H.\,A.~Roelofs,$^1$\footnote{E-mail: g.roelofs@astro.ru.nl} P.\,J.~Groot,$^1$ T.\,R.~Marsh,$^2$ D.~Steeghs$^3$, S.~Barros$^2$ \newauthor and G.~Nelemans$^{1,4}$\\
$^1$Department of Astrophysics, Radboud University, PO Box 9010, 6500 GL Nijmegen, The Netherlands\\
$^2$Department of Physics, University of Warwick, Coventry CV4 7AL, UK\\
$^3$Harvard-Smithsonian Center for Astrophysics, 60 Garden Street, Cambridge, MA 02318, USA\\
$^4$Institute of Astronomy, University of Cambridge, Madingley Road, Cambridge CB3 OHA, UK}
\begin{document}
\maketitle

\begin{abstract}
We present high time resolution VLT spectroscopy of SDSS~J124058.03$-$015919.2, a new helium-transferring binary star identified in the Sloan Digital Sky Survey. We measure an orbital period of $37.355\pm0.002$ minutes, confirming the AM CVn nature of the system. From the velocity amplitudes of the accretor and the accretion stream--disc impact, we derive a mass ratio $q=0.039\pm0.010$. Our spectral coverage extends from $\lambda\lambda$3700--9500\AA\ and shows the presence of helium, nitrogen, silicon and iron in the accretion disc, plus the redshifted, low-velocity ``central spikes'' in the helium lines, known from the low-state AM CVn stars GP Com and CE 315. Doppler tomography of the helium and silicon emission lines reveals an unusual pattern of two bright emission sites in the tomograms, instead of the usual one emission site identified with the impact of the mass stream into the accretion disc. One of the two is preferred as the conventional stream--disc impact point in velocity space, at the $3\sigma$ confidence level. We speculate briefly on the origin of the second.
\end{abstract}


\section{Introduction}

The AM CVn stars are thought to be white dwarfs accreting helium from another low-mass helium white dwarf \citep[e.g.][]{nelemans}, a semi-degenerate low-mass helium star \citep{it} or, possibly, from an initially hydrogen-rich evolved-main-sequence star \citep{podsi}. Their observed orbital periods range from 65 down to 10 or possibly even 5.3 minutes \citep{mtr, wwescet, ramsay0806, israel}. Their evolution is expected to be governed entirely by loss of angular momentum through emission of gravitational wave radiation (GWR) \citep{paczynski}. Once mass transfer has started, the mass transfer rate is determined by the loss of angular momentum from the orbit due to GWR. The mass transfer in turn determines the increase in orbital period. Since the rate of angular momentum loss due to GWR is a sensitive function of binary separation \citep{landau}, an exponential decline in mass transfer rate and orbital frequency will ensue.

The small sample of about ten known systems seems to conform to this picture rather well. It can be divided in three groups of different optical signature: (1) the short-period, high-state systems, that are believed to be in a stable state of high mass transfer. Their optical spectra are dominated by helium absorption lines, probably from the (optically thick) accretion disc; (2) the longest-period, low-state systems that are believed to be in a stable state of low mass transfer. Their optical spectra are dominated by strong helium emission lines from the (optically thin) accretion disc plus an underlying blackbody, attributed to the accreting primary white dwarf, and (3) the intermediate-period outbursting systems, whose optical signature more or less varies between that of the high and that of the low state systems, accompanied by variations in brightness of several magnitudes.

Unfortunately, it is as yet unknown which of the three suggested formation channels contribute to the AM CVn population, and in what numbers. In order to improve our overview of the population, and in particular to test its homogeneity, we started various searches for these objects. This included a search in the spectroscopic database of the Sloan Digital Sky Survey (SDSS), from which we extracted AM CVn candidates by filtering out helium emission line systems. This resulted in the discovery of SDSS J124058.03$-$015919.2 (hereafter SDSS J1240): a faint ($V=19.7$) but obvious mass-transferring binary with its characteristic strong, broad helium emission lines giving away the presence of an accretion disc \citep{roelofs}.

In addition to the helium emission, the object shows a characteristic DB white dwarf continuum with broad helium absorption lines, which fits well with a single 17,000\,K blackbody \citep{roelofs}. Recent work by Bildsten et al.\ (in preparation) predicts that the evolution of these systems towards longer orbital periods goes hand in hand with a continuous cooling of the accretor, which is ultimately due to the plummeting accretion rate. GP Com, which has an orbital period of 46 minutes, shows a continuum which is compatible with a $\sim$11,000\,K blackbody, which Bildsten et al.\ attribute to the accreting white dwarf. The accretor in the new system appears to be considerably hotter, suggestive of a higher mass transfer rate.
An inventory of previous observations of the new AM CVn star (see table \ref{previousobs}) shows that it has been observed on quite a few occasions, where it was always close to $V=19.7$, suggestive of a somewhat lower mass transfer rate than that of the longest-period outbursting AM CVn star, `SN2003aw' at $\sim$34 minutes (\citealt{ww03} and Roelofs et al.\ in prep.), although the observations do not rule out that the star exhibits outbursts just as well as `SN2003aw'.

A first attempt to measure the orbital period of the new AM CVn candidate was carried out by \citet{ww04} at the South African Astronomical Observatory through white-light photometry. However, no periodic modulation of the object's brightness was detected. A similar time-series taken independently by Ramsay \& Hakala (private communication) on the Nordic Optical Telescope gave the same results. Photometrically, SDSS J1240 thus behaves just like the two low-state AM CVn stars GP Com and CE 315 \citep{ww02}. To uncover the orbital period of SDSS J1240, as well as probe the basic kinematics of the system, we obtained time-resolved spectroscopy.

\begin{table}
\begin{center}
\begin{tabular}{l l l}
\hline
Dates		        &Survey/reference       &Type\\
                        &or observer            &\\
\hline
\hline
1979/05/19              &SSS/DSS1               &image\\
1991/05/05              &SSS/DSS2               &image\\
1999/03/12              &2QZ		        &spectrum\\
2000/03/03              &SDSS 		        &image\\
2001/03/31              &SDSS 		        &spectrum\\
2003/12/12,13           &\citet{roelofs}   &spectrum\\
2004/02/14,16,19--21    &\citet{ww04}    &photometry\\
2004/02/17              &Ramsay\,\&\,Hakala     &photometry\\
2004/12/17,18,19,30     &Roelofs                &image\\

\hline
\end{tabular}
\caption{Summary of previous (serendipitous) observations of the new AM CVn star. On all occasions, the star was close to $V=19.7$. (SSS = SuperCOSMOS Sky Survey; DSS = Digitised Sky Survey; 2QZ = 2-degree Field QSO Redshift Survey).}
\label{previousobs}
\end{center}
\end{table}

\section{Observations and data reduction}

\begin{table}
\begin{center}
\begin{tabular}{c l r r}
\hline
Date		&UT		&Exposure&	Exposures\\
                &               &time (s)&\\
\hline
\hline

FORS2/600B	&		&	&\\
2004/03/18	&06:37--07:48	&240	&16\\

FORS2/600I	&	&	&\\
2004/03/19	&07:56--08:24	&300	&6\\

FORS2/1200R	&	&	&\\
2004/04/19	&04:08--05:01	&90	&22\\
2004/04/20	&00:41--01:34	&90	&22\\
		&04:27--06:12	&90	&44\\
2004/05/10	&23:16--01:13	&90	&44\\
2004/05/11	&01:56--02:53	&90	&23\\
2004/05/13	&23:23--04:06	&90	&114\\
2004/05/17	&23:08--02:07	&90	&69\\

\hline
\end{tabular}
\caption{Summary of our VLT observations. 600B, 600I and 1200R denote the FORS2 grisms used in the set-up.}
\label{observations}
\end{center}
\end{table}

We obtained phase-resolved spectroscopy of SDSS J1240 during March, April and May of 2004 on the Very Large Telescope of the European Southern Observatory with the FORS2 spectrograph. The largest part of the observations consists of 338 high time and spectral resolution spectra in the red (grism 1200R, effective resolution $\sim$2--3\AA\ depending on seeing conditions) with the aim of carrying out Doppler tomography on the strong helium emission lines in the red, to measure beyond doubt the orbital period of the binary. We furthermore obtained a series of 16 lower-resolution spectra in the blue and 6 spectra in the far-red (grisms 600B and 600I, resolution $\sim$6\AA). Our spectral coverage thus extends from $\lambda\lambda$3800--9500\AA. A summary of the observations is given in table \ref{observations}.

All observations were done with a $1''$ slit. The detector was the FORS2 MIT CCD mosaic of which only chip 1 was used; binning was standard $2\times2$ pixels. Low read-out speed and high gain minimized the read-out and digitisation noise, respectively. In order not to add more noise to the spectra, we subtracted a constant bias level from the CCD frames. The bias level was determined per observing block of typically 22 spectra from the overscan regions on the CCD. A normalised flatfield frame was constructed from 5 incandescent lamp flatfield frames each night, which ensured a cosmic-ray-free final flatfield. For the observations taken with the 1200R grism, we used the combined flatfields taken 2004/04/19, 2004/05/10 and 2004/05/13 for all observations as suitable flatfield frames were not available for the other three nights.

All spectra were extracted using IRAF's implementation of optimal (variance-weighted) extraction. The read-out noise and photon gain, necessary for the extraction, were calculated from the bias and flatfield frames, respectively. Wavelength calibration was done with standard HeHgCd and HeNeAr arcs for the 600B and 600I spectra, respectively. A total of around 40 arc lines could be fitted well with a cubic spline of order 1 and 0.14\,\AA\ root-mean-square residual. The 1200R spectra, taken on different nights, were calibrated per observing block using 42 night sky lines, which were selected to minimize the possibilities of blending or confusion. We thus achieved a good fit on 42 sky lines, again with a cubic spline of order 1, and 0.064$\pm$0.015\,\AA\ root-mean-square residuals. All spectra were transformed to the heliocentric rest-frame prior to analysis.

The red average spectrum of SDSS J1240 was corrected for instrumental response using spectroscopic standard star EG 274. For the blue and far-red spectra, the instrumental response was removed using the flux-calibrated spectrum of SDSS J1240 from the Sloan Digital Sky Survey as a suitable flux standard was not available. In this paper, the flux calibration is mainly intended to show the continuum slope of the spectrum.

\section{Results}

\subsection{Average spectrum}

\begin{figure*}
\centering
\includegraphics[angle=270,width=\textwidth]{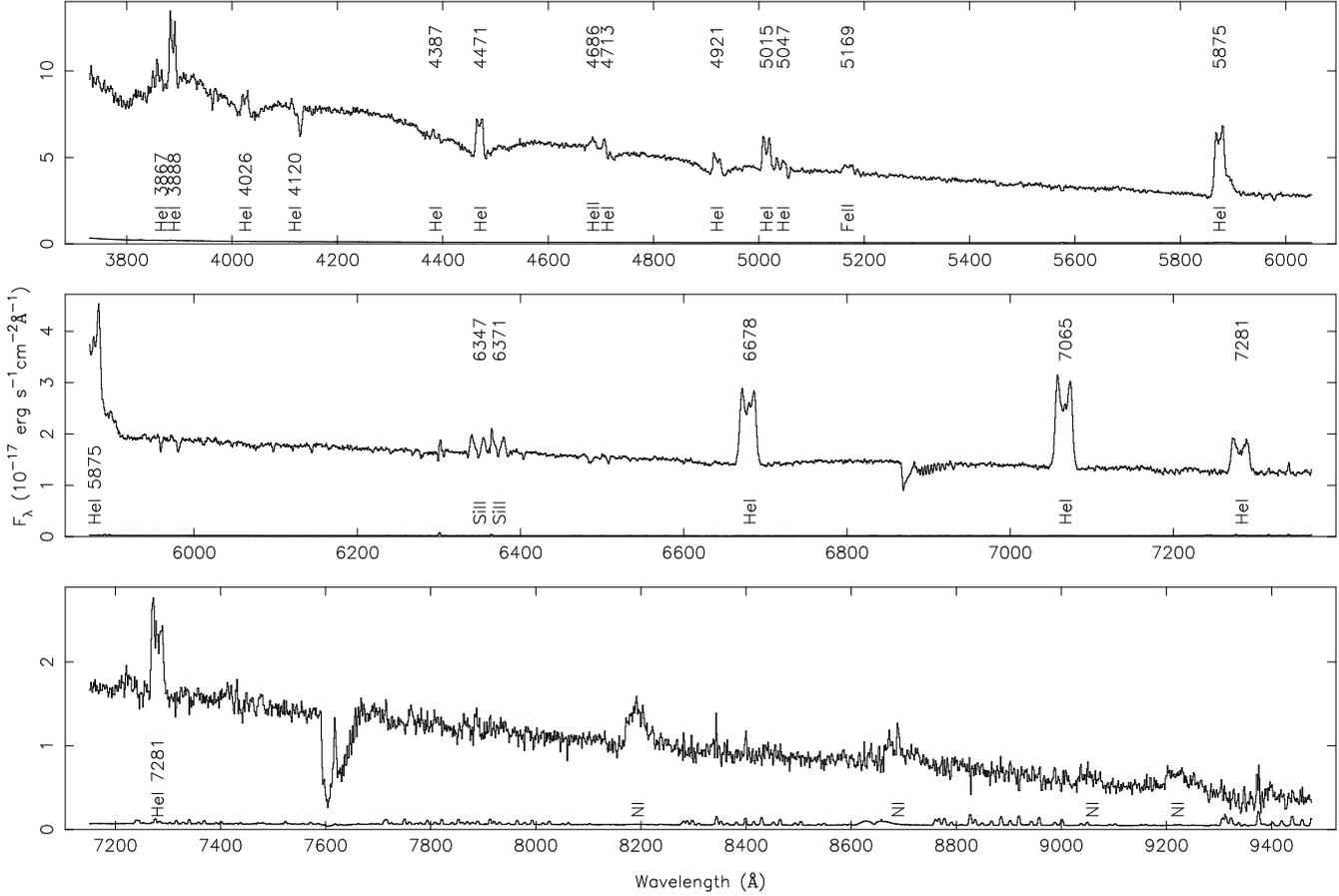}
\caption{Average spectra of SDSS J1240 with the 600B, 1200R and 600I grisms, top to bottom. The most prominent lines are labelled.}
\label{averages}
\end{figure*}

The average spectra in the blue, red and far-red are shown in figure \ref{averages}. The blue spectrum shows the familiar double-peaked helium emission lines and the aforementioned broad DB white dwarf absorption lines. In the far-red, we notice several \mbox{N\,{\sc i}} line complexes, which are quite similar to those seen in GP Com \citep[e.g.][]{lmr} and CE 315 \citep{mtr}. Both the blue and the red spectra, however, reveal features which clearly distinguish SDSS J1240 from GP Com and CE 315, namely strong \mbox{Si\,{\sc ii}} emission at $\lambda$6346\AA\ and $\lambda$6371\AA, as well as \mbox{Fe\,{\sc ii}} emission at $\lambda$5169\AA. These silicon features have also been observed in a low-state spectrum of CP Eri \citep{groot}, after model predictions by \citet{trm91} that they should show up in an optically thin helium accretion disc with solar abundances of heavy metals.

A second striking feature seen in the red spectrum, which has the largest signal-to-noise ratio and the highest spectral resolution, is the central emission in the double-winged helium emission lines (see figure \ref{doppler} in section \ref{dopplersection} for a close-up). This ``central spike'' feature was first seen in GP Com \citep{smak} and observed with the discovery of the second low-state AM CVn star, CE 315 \citep{mtr}. However, in SDSS J1240 it is considerably weaker than in the other two systems. Its origin is reported to be most likely on or very near the accreting white dwarf, based on a small but clearly detected movement of the central spike in both GP Com and CE 315, consistent in phase and amplitude with the accretor's orbital motion (\citealt{trm99}, \citealt{lmr}, Steeghs et al.\ in prep.).

\subsection{The spectroscopic period}

To find the spectroscopic period, we used a slightly modified version of the method described by \citet{nather}. The emission lines are divided into a red and blue part and the ratio of fluxes in both wings -- summed over all emission lines to maximize the signal-to-noise -- is calculated for each spectrum. In order to minimize the effects of CCD pixels falling just inside a wing mask for one spectrum, but just outside the wing mask for another due to small differences in the dispersion solution, we used masks with ``soft edges'' to calculate the red-wing and blue-wing fluxes. The edges of a mask fall off as a Gaussian which has a width matched to the spectral resolution of the set-up.

\begin{figure}
\includegraphics[angle=270,width=84mm]{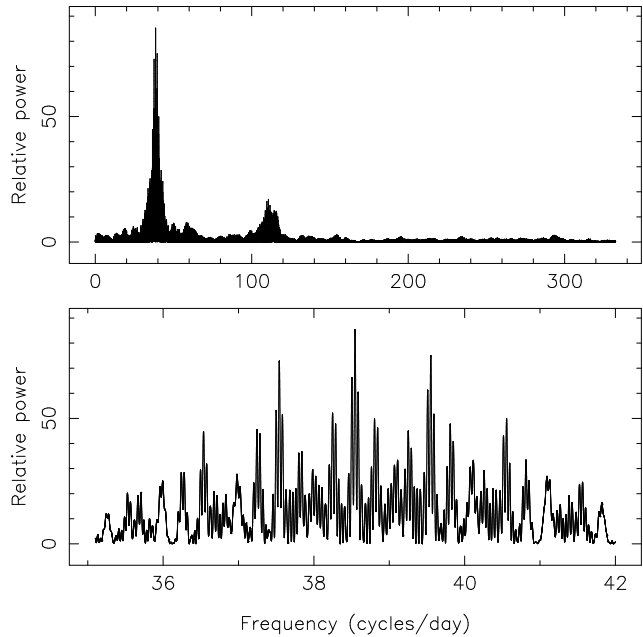}
\caption{Lomb-Scargle periodogram of the red wing/blue wing flux ratios. The lower panel provides a magnified view of the strongest peaks.}
\label{scargle}
\end{figure}

A Lomb--Scargle periodogram of the measured red-wing/blue-wing ratios as a function of heliocentric Julian date is shown in figure \ref{scargle}. A clear signal is seen at around 38 cycles/day; a weaker group of peaks appears at three times this frequency. Zooming in on the main signal, we see that the strongest peak occurs at 38.546 cycles/day ($P=37.36$ minutes), with the two next-strongest peaks occurring at the usual $\pm$1 cycle/day aliases. We will assume that the strongest peak corresponds to the orbital period of the binary. We will refine our orbital period measurement using the emission line kinematics in the next section.

\subsection{Doppler tomography}
\label{dopplersection}

\begin{figure*}
\centering
\includegraphics[angle=270,width=\textwidth]{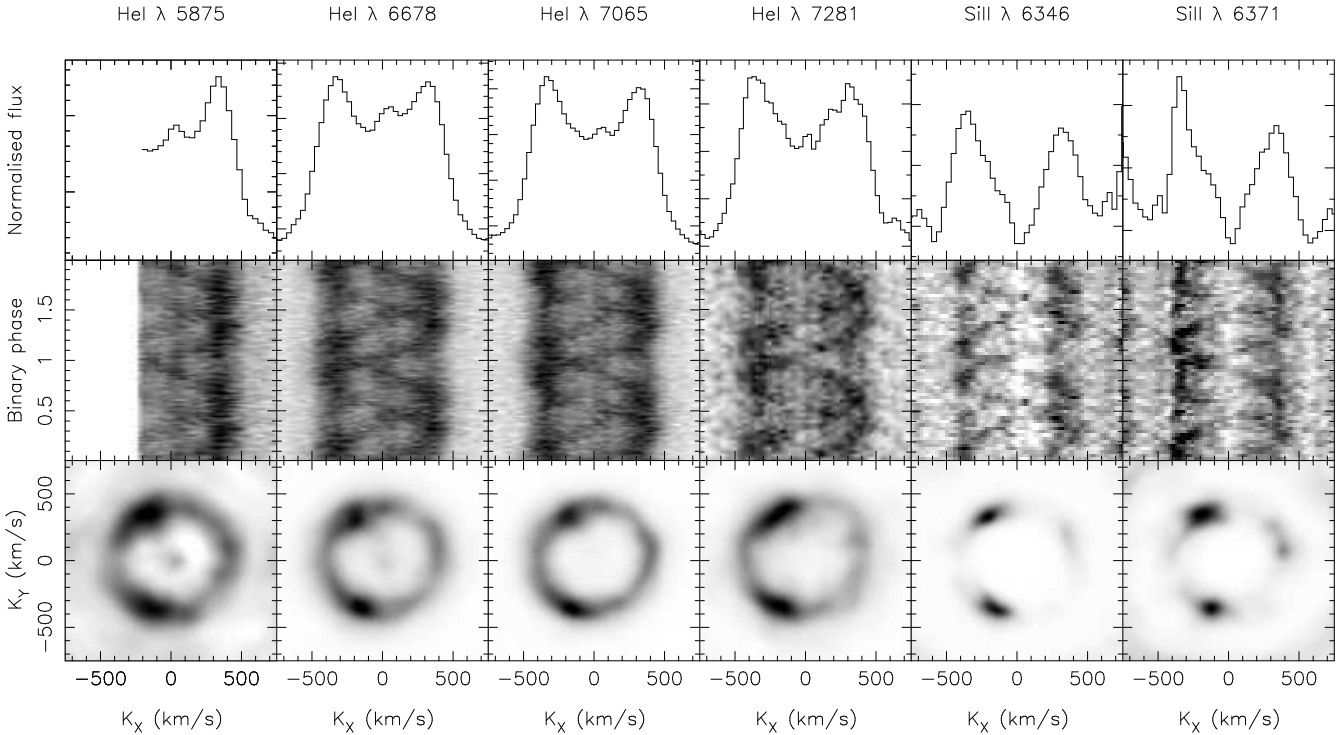}
\caption{Trailed spectra and Doppler tomograms of the R-band \mbox{He\,{\sc i}} and \mbox{Si\,{\sc ii}} features. A pattern of two bright spots, at approximately the same radial velocity but at a 120 degree angle, appears in the trails and tomograms. The central spikes are clearly visible in the trailed spectra of \mbox{He\,{\sc i}}\,5875, \mbox{He\,{\sc i}}\,6678 and \mbox{He\,{\sc i}}\,7065. The possible central spike feature in the \mbox{He\,{\sc i}}\,7281 line is strongly affected by night sky lines.}
\label{doppler}
\end{figure*}

\begin{table}
\begin{center}
\begin{tabular}{l c c c c}
\hline
Line               &$\gamma_{\mathrm{disc}}$ &$\gamma_{\mathrm{cs}}-\bar\gamma_{\mathrm{disc}}$ &$K_Y$            &$K_X$\\
                   &(km/s)                   &(km/s)                                            &(km/s)           &(km/s)\\
\hline
\hline
\mbox{He\,{\sc i}} 5876               &--                       &28.4$\pm$4.2                   &$-17.8\pm5.3$    &$5.6\pm5.3$\\
\mbox{He\,{\sc i}} 6678               &35.5$\pm$1.7             &31.5$\pm$4.2                   &$-20.8\pm5.1$    &$1.3\pm5.1$\\
\mbox{He\,{\sc i}} 7065               &33.1$\pm$1.9             &55.0$\pm$4.2                   &$-10.2\pm4.7$    &$-5.6\pm4.7$\\
\hline
\end{tabular}
\caption{Parameters for three helium lines showing a central spike: the systemic velocity as derived from the wings of the lines ($\gamma_{\mathrm{disc}}$), the intrinsic redshifts of the central spikes relative to the average systemic redshift ($\gamma_{\mathrm{cs}}-\bar\gamma_{\mathrm{disc}}$), and the velocities of the central spikes ($K_X$ and $K_Y$) for the orbital ephemeris (\ref{ephemeris}).}
\label{centralspikes}
\end{center}
\end{table}

\subsubsection{Features of the accretion disc}
\label{tomography}

Usually, a Doppler tomogram of a binary in which accretion via an accretion disc occurs, shows one bright emission site that can been attributed to the impact point of the accretion stream into the accretion disc. Both the helium and silicon emission line tomograms in SDSS J1240 show an unusual pattern of two bright spots at approximately the same radial velocity of 390$\pm$10 km/s, and at approximately the same intensity. This feature is also clearly identified in the trailed spectra. See figure \ref{doppler} for trailed spectra and maximum-entropy Doppler tomograms of all emission lines in the red.

In the trailed spectra of \mbox{He\,{\sc i}} 5875, \mbox{He\,{\sc i}} 6678 and \mbox{He\,{\sc i}} 7065 as well as in the Doppler tomograms of \mbox{He\,{\sc i}}\,5875 and \mbox{He\,{\sc i}}\,6678 we can furthermore distinguish the (rather weak) central spike, at near-zero velocities. Note that the intrinsic redshifts of the central spikes, as summarised in table \ref{centralspikes} but also clearly visible in the trailed spectra, causes them to be smeared out in the Doppler tomograms since these have been made using the rest-frame wavelengths of the helium lines. Therefore the central spikes do not show up as clearly as would be expected from the trailed spectra.

The double bright spot pattern, which persists even in Doppler tomograms made from a single night's data, was used to further refine the orbital period by aligning the bright spot patterns in the tomograms from each night. After correcting for the systemic velocity of the star, we derive an orbital period $P_\mathrm{orb}=37.355\pm0.002$ minutes. The error is determined by the time-base of our observations and the notion that a phase drift of $P_\mathrm{orb}$/20 over this time-base clearly shows up in the Doppler tomograms.

\subsubsection{Redshift and movement of the central spike}

Detailed studies of the central spike in GP Com and CE 315 \citep{trm99, lmr} have shown that it is most likely produced on or very near the accreting white dwarf. The origin of the line-dependent intrinsic redshifts observed in the central spikes of GP Com and CE 315 is still unclear and will be the subject of a separate study (Steeghs et al.\ in prep.). The main problem with the redshifts of the central spike is that it is significantly different for each helium line, which makes it difficult to explain the redshifts as being caused by just the gravitational field of the primary white dwarf. Due to the lower spectral resolution, lower signal-to-noise, more complicated line profiles, and weaker central spikes, the same approach as taken by \citet{lmr} for measuring the central spikes' redshifts in GP Com does not work for the data presented here. Instead of fitting multiple Gaussians to phase-binned spectra, we auto-correlate all spectra simultaneously via linear back-projection Doppler tomography \citep{dopplermapping}. We use the fact that, for a given trial wavelength, an emission feature in the spectra will appear blurred in a Doppler tomogram if its intrinsic ``rest'' wavelength does not coincide with the trial frequency. We thus make linear back-projection Doppler tomograms for a range of trial wavelengths around the rest-frame wavelengths of the three neutral helium lines showing a central spike, and fit a 2-D Gaussian to the central spike in every back-projection. For each line, the height (i.e., the sharpness) of the fitted central spike peaks strongly around a certain (redshifted) wavelength which we define as the ``rest'' wavelength of the central spike. The results are summarised in table \ref{centralspikes}.

Once we have the redshifts of the central spike, we can look for the location of the central spikes in the velocity space of their Doppler tomograms, which should reflect the orbital motion of the primary white dwarf if we assume that the central spike is indeed produced on or near the primary white dwarf. Measuring the central spike's movement thus allows us to (a) determine the zero phase in the orbital ephemeris, and (b) constrain the mass ratio of the system from the ratio of velocities of the central spike and the bright spot. To further optimise the signal-to-noise we fit a 2-D Gaussian to the combined back-projections of all three helium lines showing a central spike, where the central spikes' centre wavelengths that we just determined are taken as the ``rest'' (zero-velocity) wavelengths.

Fitting a 2-D Gaussian to the central spike in a back-projection requires separating the low-velocity spike from the higher-velocity asymmetric accretion disc emission, which may cause gradients that shift the peak of the central spike. We assume that the background gradients in the central spike region of the back-projection are caused by high-velocity emission sites (e.g., the bright spots) that blurr into the low-velocity spike region, such that we can estimate the background gradients at radial velocities below, say, 100 km/s by considering the intensity at the $\sqrt{K_X^2 + K_Y^2} = 100$\,km/s boundary. We then interpolate the background underneath the central spike by solving Laplace's equation for the intensities of the pixels in the inner region of the back-projection given the intensities (``potential'') along the boundary.

The exact choice of boundary is arbitrary; reasonable values are 100--150 km/s, given the central spike's width of about 40 km/s and the strong accretion disc emission peaking around 400 km/s. When determining the central spikes' redshifts, we varied the boundary radius between 100--150 km/s to estimate the uncertainties on the redshifts as shown in table \ref{centralspikes}.

The uncertainty on the central spike's velocity as obtained from the 2-D Gaussian fit is determined with a simple Monte Carlo simulation. We follow the procedure above for a large number of Doppler tomograms in which we set random phases on all spectra. One then expects the central spike to be smeared out (slightly) around $K_X = K_Y = 0$ in each random tomogram. A 2-D Gaussian fit to such a randomised central spike should thus lie at $K_X = K_Y = 0$; the deviations from this expected value give us an estimate of the uncertainty in our measurement. After fitting all the randomised central spikes, we fit a 2-D Gaussian to the distribution of scatter around $K_X = K_Y = 0$. This fits well with $\chi^2/\mathrm{n.d.f}\approx 1$. The radial width of this Gaussian is taken as the 1-$\sigma$ accuracy of our central spike velocity measurement. The results for the central spike and its corresponding error turn out to be insensitive to the exact choice of the background boundary radius in the range 100--150 km/s. For radii smaller than 100 km/s and for radii larger than 150 km/s, the scatter in the fits to the random tomograms increases.

Using this method, we obtain a central spike velocity of $15.3 \pm 4.6$\,km/s. By aligning the central spike along the negative $K_Y$ axis, which is where the primary white dwarf should be, we obtain the zero phase in the ephemeris of SDSS J1240. With the spectroscopic period found above, the full orbital ephemeris then becomes
\begin{equation}
T_0(\mathrm{HJD}) = 2453115.6599 + 0.025944 E
\label{ephemeris}
\end{equation}
As a further sanity check on the significance of the determined central spike velocity, we repeat the central spike velocity measurement for each of the helium lines separately. Although the individual velocity measurements will have larger error than the measurement above from the three lines combined, agreement between the individual lines on the central spike velocities would give strong support to the correctness and significance of the determined amplitude and phase.

Table \ref{centralspikes} shows the results for the spike velocities for each line after employing the orbital ephemeris (\ref{ephemeris}). Indeed the individual lines yield mutually compatible locations for the central spike in velocity space, along the negative $K_Y$ axis, giving support to the statistical significance of the determined zero phase in the ephemeris (\ref{ephemeris}). Figure \ref{doppler} shows the maximum-entropy Doppler tomograms \citep{dopplermapping} that we obtain from this orbital ephemeris. We see that one of the two bright spots lines up near the expected stream--disc impact point in velocity space, especially for ballistic stream velocities at the impact site (cf.\ figure \ref{brightspots}), while the other bright spot ends up in the lower-left quadrant. The uncertainty in the zero phase is on the order of $\arcsin \left(4.6/15.3\right)\approx 20^{\circ}$ or 0.0013 days.

\subsubsection{System parameters}

From the low projected velocity amplitude of the primary ($v_\mathrm{1,proj}=15.3\pm4.6$ km/s) and the projected velocity amplitude of the bright spot ($v_\mathrm{spot,proj}=390\pm10$ km/s) we can put interesting limits on the mass ratio $q = M_2 / M_1$ of the system. Since we have two velocity amplitudes, the unknown orbital inclination is eliminated quite trivially. The independence of the velocity ratio on the absolute mass of the components is slightly less trivial, but is easily demonstrated (appendix \ref{systempars}). We are left with just two parameters that determine the velocity ratio of the primary and the bright spot: the mass ratio $q$, and the effective accretion disc radius $R$ at the stream--disc impact point.

\begin{figure}
\includegraphics[angle=270,width=84mm]{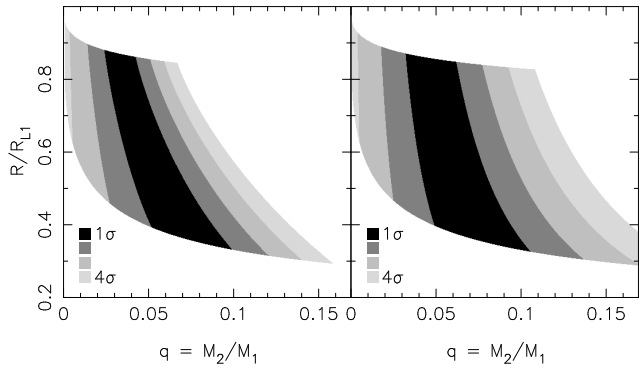}
\caption{Allowed values of the accretion disc radius $R$ and the mass ratio $q$ for SDSS J1240. Left the results if one assumes ballistic stream velocities in the bright spot. In the right panel the results in case of disc velocities.}
\label{limits}
\end{figure}

Figure \ref{limits} shows the limits on $q$ and $R$ obtained by solving the equation of motion for a ballistic particle in the accretion stream (\ref{equationmotion}) for a grid of values for $q$ and $R$. A hard upper limit on $R$ is caused by the accretion disc extending beyond the Roche lobe for large $R$. Similarly, a hard lower limit on $R$ is set by conservation of angular momentum of the matter falling in from the inner Lagrange point, and the resulting constraint that the accretion disc should \emph{at least} extend to the circularisation radius $R_\mathrm{circ}$, the radius at which the specific angular momentum for matter in a Keplerian orbit around the primary equals that of matter in the $L_1$ point. In the remaining parameter space, a greyscale plot indicates the regions allowed by the data at various confidence levels.

We see that, for reasonable disc radii of $0.7-0.8 R_{L_1}$ and assuming ballistic stream velocities in the bright spot, we find a mass ratio of about $q=0.039\pm0.010$.\footnote{$R_{L_1}$ represents the distance from the centre of the primary to the $L_1$ point, not the effective radius of the primary Roche lobe.} In reality, there may be mixing with accretion disc velocities in the bright spot region. In case of pure Keplerian disc velocities the mass ratio increases slightly to $0.050\pm0.015$. We can conclude that the system is indeed of extreme mass ratio, as expected for an AM CVn star with a 37-minute orbital period. The mass ratio falls between $q \approx 0.087$ for AM CVn itself at $P_\mathrm{orb}=1028$\,s \citep{nsg}, and $0.017<q<0.020$ for GP Com at $P_\mathrm{orb}=2970$\,s (Steeghs et al.\ in prep.).

\section{Discussion}

\subsection{System parameters}

We presented high time resolution optical spectra of SDSS J1240 and determined an orbital period of 37.355$\pm$0.002 minutes, confirming beyond doubt the AM CVn nature of the system. This is further supported by the extreme derived mass ratio of 0.039$\pm$0.010.

It is interesting to compare the new system with theoretical predictions. Bildsten et al.\ (in prep.) have recently modelled the temperature of the accretors in AM CVn stars as a function of orbital period. Depending on the core temperature and mass of the donor star, they find an accretor temperature between 11,000--20,000\,K for an AM CVn star with a 37-minute orbital period. The temperature of about 17,000\,K derived for SDSS J1240 \citep{roelofs} is consistent with that prediction. With a more accurate measurement of the primary temperature, it may be possible to constrain the entropy of the donor star for SDSS J1240 based on the results of Bildsten et al., since the higher its entropy, the higher the mass transfer rate, and the hotter the primary for any given orbital period. High signal-to-noise flux-calibrated spectra in the UV would be needed for this, combined with realistic DB white dwarf atmosphere models.

We can estimate the distance to SDSS J1240 with the absolute magnitude $M_V \simeq 11.5$--12.0 as modelled by Bildsten et al., which is set by the effective temperature and radius of the accreting white dwarf. This gives a distance of $350-440$ pc and a height $z=305-385$ pc in the Galactic disc.

For a given mass--radius relation of the Roche-lobe filling donor star, we can calculate the component masses using the measured orbital period and mass ratio. If we assume a fully degenerate, zero-temperature helium white dwarf (\citealt{zapolsky}, \citealt{rappaport}) we get a secondary mass of $0.012M_{\sun}$, which means a primary mass of $0.31^{+0.10}_{-0.07}M_{\sun}$ and an inclination $i\sim53^{\circ}$. These are the minimum allowed masses for the system. For a semi-degenerate helium star secondary \citep[see][]{nelemans} we obtain $M_2=0.031M_{\sun}$, $M_1=0.79^{+0.28}_{-0.16}M_{\sun}$ and $i\sim36^{\circ}$.

\subsection{Chemical composition and temperature of the accretion disc}

The presence of \mbox{Si\,{\sc ii}} and \mbox{Fe\,{\sc ii}} clearly distinguishes SDSS J1240 from GP Com and CE 315. \mbox{Si\,{\sc ii}} was first observed in an AM CVn star by \citet{groot} in a quiescent spectrum of CP Eri. A decade earlier, \citet{trm91} noticed the strong underabundance of Si and Fe in GP Com (less than 1/1000 solar) based on the fact that they did \emph{not} show up in the spectrum \emph{at all}, and concluded that GP Com is probably a population II halo object. The identical spectral signature of CE 315, combined with its high proper motion, suggests a similar origin for CE 315. The spectrum of the new star SDSS J1240, like that of CP Eri, indicates a higher metallicity. A simple LTE model of a helium-dominated disc of about 11,000\,K with solar abundances of heavy metals (similar to that used by \citet{trm91}, see \citet{nelemans04} for a description) predicts that \mbox{Si\,{\sc ii}} 6346 \& 6371, \mbox{Fe\,{\sc ii}} 5169 and \mbox{Ca\,{\sc ii}} H \& K should be the strongest metal lines. Apart from the calcium lines, which are not detected, this agrees with our observations. Furthermore, although the model is undoubtedly too simple, the strengths of the metal lines relative to helium indicate that the abundances of silicon and iron are compatible with solar values. We therefore expect that SDSS J1240, unlike GP Com and CE 315, is a more ``ordinary'' population I object. This is further supported by the low proper motion $\mu=18\pm17$ mas/yr of the object, as measured by SuperCOSMOS from scans of southern sky survey plates.

Interestingly, a blue spectrum of `SN2003aw' in its low state looks identical to that of SDSS J1240 shown in figure \ref{averages} but for the presence of strong \mbox{Ca\,{\sc ii}} H \& K emission (Roelofs et al.\ in prep.). This is further indication that SDSS J1240, or at least the material that we are seeing now in the disc, may somehow be underabundant in calcium.

\subsection{The second bright spot}

\begin{figure}
\includegraphics[angle=270,width=84mm]{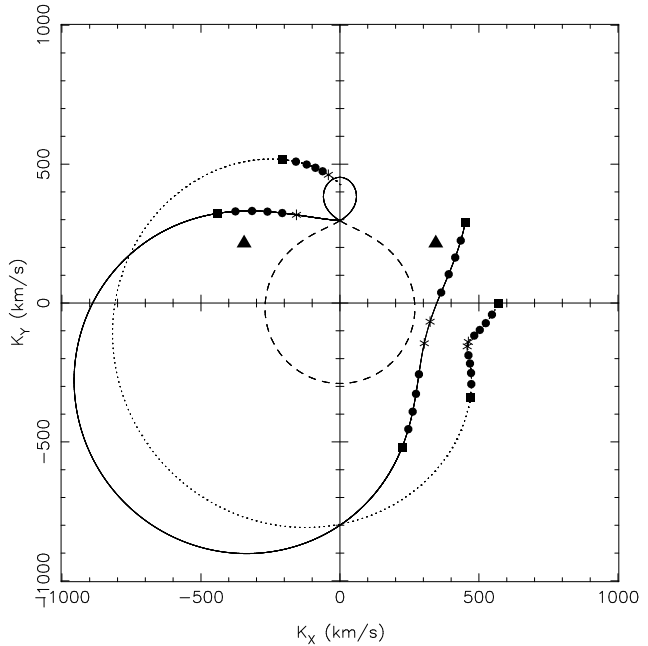}
\caption{Possible scenarios for the second bright spot for a $q=0.04$ binary. The solid line represents the stream velocities along the ballistic stream starting at the inner Lagrange point, while the dotted line indicates the Keplerian disc velocities at each point along the path followed by the stream. Adjacent spots represent the impact points for disc radii from $R=0.6R_{L_1}$ (squares) to the maximum disc radius that can be contained within the primary Roche lobe (asterisks). The group of spots at $K_X > 0$ are the second and third crossing of the accretion disc edge by the ballistic stream. The triangles indicate the $L_4$ and $L_5$ points.}
\label{brightspots}
\end{figure}

When determining the mass ratio of the system, we assumed that one of the bright spots corresponds to the (first) stream--disc impact point. This is supported by the fact that one of the bright spots lines up with the expected stream--disc impact position in the tomograms if we use the zero-phase as determined by the central spike.

The origin of the second bright spot in the trailed spectra and Doppler tomograms is as yet unclear. Given the fact that the radial velocity of both bright spots is identical, one would expect a similar origin. The first thing that comes to mind is an overshoot scenario in which part of the accretion stream initially ``misses'' the (flared) edge of the accretion disc, and impacts the disc rim again on the other side of the disc. However, the stream would have to deviate significantly from a ballistic trajectory after the first impact point in order to produce the spot at the angle observed, in which case one would expect the radial velocity of the second bright spot to change significantly as well, contrary to what is observed. Figure \ref{brightspots} shows the ballistic trajectory and its second and third encounters with the edge of the accretion disc for a wide range of disc radii; clearly, no bright spot occurs in the lower-left quadrant of the Doppler tomogram.
Another possibility is that the particle orbits in the outer disc get excited by the impact, causing a standing wave pattern in the outer disc, while a third possibility would be that the second bright spot is not caused by the mass stream, but is simply a tidal effect of the secondary and the disc, like the spiral arms observed in outbursting dwarf novae \citep{steeghsspiral}.

If one considers the 120-degree angle between the bright spots as a starting point, then the ``Trojan asteroids'' at the $L_4$ and $L_5$ points come to mind as a natural 120-degree angle in any binary system. Furthermore, the extreme derived mass ratio $q = 0.039\pm 0.010$ coincides with the stability criterion for matter in the $L_4$ and $L_5$ point ($q < 0.040$). However, as can again be seen in figure \ref{brightspots}, the positions of the two spots in the Doppler tomograms as determined from the phase of the central spike do not match the expected $L_4$ and $L_5$ positions.

It should be stressed that the fact that the double bright spot pattern persists even in data from a single night rules out the possibility that they are the result of a mistake in identifying the correct orbital period alias. The double bright spot feature thus appears to be real. The correctness or statistical significance of our determined zero-phase is more open to debate, in which case the entire Doppler tomogram could be freely rotated around $K_X = K_Y = 0$, opening up more possibilities for the bright spots' origin.

Interestingly, VLT+UVES spectroscopy of both GP Com and CE 315 (Steeghs et al.\ in prep.) shows a very weak second bright spot at approximately the same place as in SDSS J1240. In these high-resolution data there is no room for a different zero-phase. Although the second bright spot is much weaker relative to the `primary' bright spot than in SDSS J1240, it does suggest that the second bright spot feature, like the enigmatic central spike, may be another peculiar feature common to (and unique to) the AM CVn stars.

\section{Acknowledgments}

GR and PG are supported by NWO VIDI grant 639.042.201 to P.J. Groot. DS acknowledges a Smithsonian Astrophysical Observatory Clay Fellowship. GN is supported by NWO VENI grant 639.041.405 to G. Nelemans. TRM is supported by a PPARC Senior Research Fellowship. We thank G. Ramsay and P. Hakala for sharing their results on photometry of SDSS J1240. This work is based on data taken at the European Southern Observatory, Chile, under programmes 073.D-0631(A) and 072.D-0052(A).

\appendix

\section{The system parameters}
\label{systempars}

Consider a binary system in which all velocity vectors are in the orbital plane. The ratio of the projected bright spot velocity and the projected primary star velocity in a Doppler map is then given by
\begin{equation}
\frac{v_\mathrm{spot,proj}}{v_\mathrm{1,proj}} = \frac{|{\bf v}_\mathrm{spot} + \omega \times {\bf r}_\mathrm{spot}|}{|\omega \times {\bf r}_1|}
\label{ratio}
\end{equation}
To demonstrate the independence of this ratio on the absolute masses of the binary components, it will suffice to show that (i) the lengths of ${\bf r}_1$, ${\bf r}_\mathrm{spot}$ and ${\bf v}_\mathrm{spot}$ all scale with the same function of $M_1$, and that (ii) the angle between ${\bf r}_\mathrm{spot}$ and ${\bf v}_\mathrm{spot}$ remains the same.
Consider the Roche potential
\begin{equation}
\Phi({\bf r}) = -\frac{G M_1}{|{\bf r}-{\bf r}_1|} - \frac{G M_2}{|{\bf r}-{\bf r}_2|} - \frac{1}{2}({\bf \omega} \times {\bf r})^2
\label{roche}
\end{equation}
where ${\bf r}$ points from the centre of mass, ${\bf r}_{1,2}$ denote the positions of the primary and secondary star, respectively, and the angular frequency $\omega$ is related to the masses and binary separation $a$ as
\begin{equation}
\omega = \left[\frac{G\left(M_1 + M_2\right)}{a^3}\right]^{1/2}
\label{omega}
\end{equation}
If we scale up the system by increasing the masses of the components while keeping their ratio fixed, it is clear from (\ref{roche}) and (\ref{omega}) that the potential difference between two points, say the inner Lagrange point and a point fixed on the edge of the accretion disc, scales as
\begin{equation}
\Delta\Phi \propto \frac{M_1}{a}
\end{equation}
Therefore the speed gained by a ballistic particle (in the binary frame) goes as $v_{\mathrm{bal}}\propto\sqrt{M_1/a}$, while in a Keplerian disc, the velocities of matter orbiting the primary also scale as $v_{\mathrm{kep}}\propto\sqrt{M_1/a}$. Therefore, both in case of ballistic stream and Keplerian disc velocities in the bright spot, we have
\begin{equation}
|{\bf v}_{\mathrm{spot}}| \propto \sqrt{\frac{M_1}{a}}
\label{proportional1}
\end{equation}
while from (\ref{omega}) one sees that for all ${\bf r}$,
\begin{equation}
|\omega \times {\bf r}| \propto \sqrt{\frac{M_1}{a}}
\label{proportional2}
\end{equation}
Assuming an ideal Roche geometry, the full equation of motion for a ballistic particle becomes
\begin{equation}
\frac{\partial^2 {\bf r}}{\partial t^2} = - \nabla \Phi - 2\left(\omega \times {\bf v}\right)
\label{equationmotion}
\end{equation}
in which it is now easy to verify that both terms on the right hand side scale as $M/a^2$. Hence, the shape of the ballistic stream is independent of the absolute mass scale. The angle between ${\bf v}_{\mathrm{spot}}$ and ${\bf r}_{\mathrm{spot}}$ will thus remain constant. Combined with (\ref{proportional1}) and (\ref{proportional2}) we may therefore conclude that the ratio of the projected bright spot and primary star velocities (\ref{ratio}) is indeed independent of the absolute masses.

\end{document}